\documentclass{elsart}
\usepackage{graphicx}
\usepackage{amssymb}
\journal{Nuclear Physics A}

\begin{document}
\begin{frontmatter}
\title{Using Continuum Level Density in the Pairing Hamiltonian: BCS and Exact Solutions}
\author{R. Id Betan}
\address{Department of Physics and Chemisty (FCEIA-UNR) -
         Physics Institute of Rosario (CONICET),
         Av. Pellegrini 250, S2000BTP Rosario, Argentina}

\begin{abstract}
Pairing plays a central role in nuclear systems. The simplest model for the pairing is the constant-pairing Hamiltonian. 
The aim  of the present paper is to include the continuum single particle level density in the constant pairing Hamiltonian and to
make a comparison between the approximate BCS and the exact Richardson solutions.
The continuum is introduced by using the continuum single particle level density. 
It is shown that the continuum makes an important contribution to the pairing parameter even in those case when the continuum is weakly populated.
It is shown that while the approximate BCS solution depends on the model space the exact Richardson solution does not.
\end{abstract}

  
\end{frontmatter}

\section{Introduction} \label{sec.introduction}
Many-body calculations generally start with a mean-field which provides the single-particle representation from which the many-body representation is built.
The preferred framework for studying nuclear many-body systems is the interacting Shell Model (ISM) and its extensions to open systems using real \cite{69Mahaux,03prOkolowicz,06prcVolya} or complex energy \cite{02prlIdBetan,02prlMichel,09jpgMichel} representations.
The drawback of the ISM is that it becomes cumbersome as the dimension of the model space or the valence particles increase. The situation becomes even worse when the continuum part of the energy spectrum is included.
Even when the pairing Hamiltonian includes much less correlations than the ISM it contrives to get the most important part of the pairing interaction. The pairing Hamiltonian constitutes an important approximation in nuclear systems \cite{06psSatula,05Brink,2011Potel} and it can be considered as a first approximation to a large-scale continuum shell model calculation.

The pairing Hamiltonian can be solved in an approximate way by using the Bardeen-Cooper-Schrieffer (BCS) approximation. In references \cite{97plbSandulescu,01prcKruppa,07npaDussel} the BCS solution was studied by using a single particle basis which includes the complex energy continuum as well.
However the BCS solution has two main drawbacks: (i) the many-body wave function does not conserve the number of particles and (ii) it does not produces a non-trivial solution for those values of the strength which are smaller than a critical value. The constant-pairing Hamiltonian can have exact solution \cite{64npRichardson} if we use the similarity between the many-body time-reversed pairs with those of the many-body boson systems. The exact solution does conserve the particle number, moreover the solution exits for all values of the pairing strength. The eigenfunctions and eigenvalues are written in terms of a set of parameters called pair energies. In Ref. \cite{03prcHasegawa} the Richardson solution was used in a representation which included the resonant part of the continuum. 

In this work an exact solution of the pairing Hamiltonian is given in a representation which includes the continuum in a real energy representation.
The continuum is included through the continuum single particle level density (CSPLD). In order to eliminate the spurious effects of the so called particle gas \cite{73plbMosel,05prcCharity}, the CSPLD relative to a free particle in a box \cite{37pBeth} is used. 
In order to asses the limitations of the BCS solution
the  CSPLD is introduced into the approximate BCS solution.

In Section \ref{sec.method}  
the pairing Hamiltonian with a continuum basis and the expression of the BCS and Richardson equations are given. The application of the methods for the Sn isotopes are presented in Section \ref{sec.result}. The conclusion and future perspective are summarized in  Section \ref{sec.discussion}.

\section{Method}\label{sec.method}
\subsection{Hamiltonian}
The pairing-model Hamiltonian in a continuum basis reads,
\begin{equation}
 H = H_{sp} - G \; P^+ \; P
\end{equation}

with the operators
\begin{eqnarray}
 H_{sp} &=& \sum_\beta \varepsilon_b \; a^+_\beta a_\beta + 
            \int_0^\infty d\varepsilon \; \varepsilon\; 
                   \sum_\gamma a^+_\gamma(\varepsilon)a_\gamma(\varepsilon) \nonumber \\
 P^+ &=& \sum_{\beta>0} \; a^+_\beta a^+_{\bar{\beta}} + 
         \int_0^\infty d\varepsilon \sum_{\gamma>0} a^+_\gamma(\varepsilon) a^+_{\bar{\gamma}}(\varepsilon) \nonumber 
\end{eqnarray}

Here the index $\beta=\{ b,m_\beta \}=(n_b,l_b,j_b,m_\beta)$ refers to bound states and $\gamma=\{ c, m_\gamma \}=(l_c,j_c,m_\gamma)$ to continuum states. The first summation is over the valence bound states while the second one is over the partial waves. In practical applications upper limits are set for the energy $\varepsilon_{max}$ and for the partial waves $l_{max}$ in which continua are taken into account. These parameters determine the size of the selected model space.
The creation $a^\dagger$ and annihilation $a$ operators satisfy the usual anti-commutation relationship with Kronecker  delta for the bound states and Dirac delta for the continuum states.
The operator $a^+_{\bar{\nu}}=(-)^{j_n-m_\nu} a^+_{n -m_\nu}$ is the time reversed of the $a^+_\nu$ operator. The summation $\nu > 0$ refers to only the positive values of the $m_\nu$, i.e. projection of  the total angular momentum.
It is assumed that the pairing strength parameter $G$ parameterize the total number of particles ($A=A_{core}+A_{valence}$) as $G=\chi/A$ \cite{63rmpKisslinger}.
For a given model space the parameter $\chi$ will be adjusted to reproduce the pairing energy taken from theoretical mass table for the isotope of the middle of the shell. 

\subsection{Model Space: Bound and Continuum States}
The single particle model space is calculated in a Woods-Saxon (WS) potential
describing the mean field. For simplicity the energy shift of the levels due to the change of the atomic mass number will be ignored in this work, i.e. the same single particle energies will be used for all isotopes. The mean-field parameters are chosen to reproduce approximately the experimental energies of the core plus one nucleon system. The continuum part of the spectrum is represented by the continuum single particle level density (CSPLD). The CSPLD is defined relative to the free particle density in terms of the phase shift $\delta$ \cite{37pBeth,73plbMosel,78plbFowler,79plbTubbs,84npaBonche,85zpDean,92npaSholmo,96prcDobaczewski,98plbKruppa,05prcCharity}. 
The phase shifts are obtained by solving the time independent Schroedinger equation for positive real energy $\varepsilon$ \cite{82Newton} in the WS potential for calculating bound states of the valence particles  

\begin{equation} \label{eq.ge}
 g(\varepsilon) = \sum_c \frac{2j_c+1}{\pi} \frac{d\delta_c}{d\varepsilon}
\end{equation}

where $\delta_c(\varepsilon)$ is uniquely determined by the requirement of its continuity in the energy region concerned.

\subsection{BCS Equations}
The gap parameter $\Delta$ in the continuum basis reads \cite{97plbSandulescu},
\begin{eqnarray}
 \Delta &=& \Delta_b + \Delta_c  \nonumber \\
 \Delta_b &=& \frac{G}{2} \sum_b (2j_b+1) u_b v_b  \label{eq.gap-b} \\
 \Delta_c &=& \frac{G}{2} \int_0^\infty d\varepsilon\;
               u(\varepsilon) v(\varepsilon) g(\varepsilon)  \label{eq.gap-c}
\end{eqnarray}

where $u_b$ and $u(\varepsilon)$ are the usual occupation probabilities amplitude for bound and continuum states respectively in the Biedenharn-Rose phase convention \cite{2007Suhonen}.

The gap and the particle number equations are the following
\begin{eqnarray} 
 \frac{4}{G} &=& \sum_b \frac{(2j_b+1)}{E_b} 
                  + \int_0^\infty d\varepsilon\; \frac{g(\varepsilon)}{E(\varepsilon)}~, \\
 N &=& N_b + N_c \\
 N_b &=& \sum_b (2j_b+1) v^2_b  \label{eq.nd}\\
 N_c &=& \int^\infty_0 \;d\varepsilon \; v^2(\varepsilon)\; g(\varepsilon) \label{eq.nc}
\end{eqnarray}

where $E_b$ and $E(\varepsilon)$ are the usual quasi-particle energies in the bound and continuum states respectively \cite{2007Suhonen}. The CSPLD modifies the pairing strength in the gap equation in a effective way  and it prevents us from putting non-physically large numbers of particles into the continuum.

The ground state energy $E_{BCS}$ reads,
\begin{equation}\label{eq.bcs}
 E_{BCS} = \sum_b (2j_b+1) v_b^2 \left( \epsilon_b - \frac{G}{2} v_b^2 \right) 
         + \int^\infty_0 \;d\varepsilon \; v^2(\varepsilon)\; g(\varepsilon) 
           \left[ \varepsilon - \frac{G}{2} v^2(\varepsilon) \right] 
         - \frac{\Delta^2}{G}
\end{equation}

\subsection{Richardson Equations}
A feasible generalization of the Richardson equations to basis with a continuum part can be found in Ref. \cite{2004Dukelsky},
\begin{equation} \label{eq.r}
 1 - \frac{G}{2} \sum_b \frac{2j_b+1}{2 \varepsilon_b - E_\alpha} 
   - \frac{G}{2} \int_0^\infty d\varepsilon  \; \frac{g(\varepsilon)}{2 \varepsilon - E_\alpha} 
   + 2 G \sum_{\beta \ne \alpha} \; \frac{1}{E_\beta - E_\alpha} = 0
\end{equation}

where $\varepsilon_b$ and $E_\alpha$ are the bound single particle  energies in the mean-field  and the pair energies \cite{63plRichardson}, respectively.
The CSPLD is included in the coupled equations which determines the pair energies. The number of equations is equal to the number of pairs in the valence configuration \cite{63plRichardson,64npRichardson}. If we make an analytic continuation of Eq. (\ref{eq.r}) to the complex energy plane it can be reduced to the form of ref. \cite{03prcHasegawa}.

The ground state energy written in terms of the pair energies is \cite{63plRichardson},
\begin{equation} \label{eq.rich}
 E_{Rich} = \sum_\alpha \; E_\alpha
\end{equation}

\subsection{Correlated energy}
The pair-correlation energy is defined by the difference between the energies with and without pairing. The energies with pair correlations in the BCS approximation and in the Richardson model are given in Eq. (\ref{eq.bcs}) and in Eq.(\ref{eq.rich}), respectively. While the energy without pair correlation in the continuum basis reads,
\begin{equation}
 E_0 = \sum_b n^0_b \left( \epsilon_b - \frac{G}{2} \right) 
         + \int^\lambda_0 \;d\varepsilon \; \theta(\lambda-\varepsilon)\; g(\varepsilon) 
           \left( \varepsilon - \frac{G}{2} \right)
\end{equation}

The occupation number $n^0_b$ is the number of particles in the level $b$ below the Fermi level  $\lambda$ and it is zero above $\lambda$. The step function $\theta(\lambda-\varepsilon)$  makes the integral act only in those cases when the Fermi level is in the continuum.

The correlation energy reads,
\begin{equation} \label{eq.ecorr}
  E_{corr} = E_{model} - E_0~,
\end{equation}
where model denotes BCS Eq. (\ref{eq.bcs}) or Richardson Eq. (\ref{eq.rich}).

\subsection{Pairing Strength}
The strength parameter $\chi$ in $G=\frac{\chi}{A}$, depends on the size of the 'model space' and naturally on the 'model solution' too. The model space is define by the cut-off energy $\varepsilon_{max}$ and the maximum orbital angular momentum $l_{max}$, while model solution refers to the approximate BCS solution or to the exact Richardson solution. 
For a given model space the strength parameter $\chi$ is chosen to reproduce the experimental pairing energy calculated from mass table for one of the isotopes concerned and it is keep constant for all the other isotopes.

In order to determine the relationship between the experimental pairing energy $P$ and the model-solution pairing energy $P_{model}$ we will use the expression in terms of the ground-state energies $E$ \cite{2007Suhonen} and we require $P=P_{model}$. To be specific, let us assume that the proton states form a single close shell and we have an even number $N=2N_{pair}$ of neutrons,
\begin{equation}
 P(N)= 2E(N-1)-E(N)-E(N-2)
\end{equation}

For the BCS model-solution $P_{BCS}=\Delta$ \cite{2007Suhonen} while for the Richardson model-solution \cite{64npRichardson}
\begin{equation}
 P_{Rich} = 2 \varepsilon_{p_{N_{pair}}} - Re\left[ E_{p_{N_{pair}}}(2N_{pair}) \right]
\end{equation}
In order to explain the meaning of the indexes in the single particle energy $\varepsilon_{p_i}$ and in the pair energy $E_{p_i}$ let us consider an example. Let us use the following $M_b$ model space: $\{1f_{7/2},\;2p_{3/2},\;0h_{9/2},\;2p_{1/2},\;1f_{5/2}\}$ and $N_{pair}=8$, i.e. there are $N=16$ valence particles. We can label the $16$ single particle energies in the ground state as $(\varepsilon_1,\;\varepsilon_2,\;\dots,\;\varepsilon_{16})$ or as $(\varepsilon_{p_1},\;\varepsilon_{p_2},\;\dots,\;\varepsilon_{p_8})$. For the $M_b$ model space $\varepsilon_{p_1}=\dots=\varepsilon_{p_4}=\varepsilon_{1f_{7/2}},\;\varepsilon_{p_5}=\varepsilon_{p_6}=\varepsilon_{2p_{3/2}},\;\varepsilon_{p_7}=\varepsilon_{p_8}=\varepsilon_{0h_{9/2}}$. 
Then we define the pair energy by the limit $lim_{G\rightarrow0^+} E_{p_i}=2\varepsilon_{p_i}$. 

The strength parameter $\chi$ is fixed for a given isotope $N$ for each model-space and model-solution. For the BCS  $\chi$ is chosen by the requirement
 $P(N)=\Delta$. For the Richardson model-solution $\chi$ is adjusted to reproduce the energy of the last pair $Re[E_{p_{N_{pair}}}(2N_{pair})]= 2\varepsilon_{p_{N_{pair}}}-P(N)$.

\section{Application} \label{sec.result}
The method outlined in secc. \ref{sec.method} will be applied to the Tin isotopes above the $^{132}Sn$ core.

\subsection{Setting the Parameters}
\subsubsection{Model Space: Bound and Continuum States}
The mean-field which describes the valence bound states and the continuum states is represented by a Woods-Saxon plus spin-orbit potentials with the following parameters: $V_0=43.5$ MeV, $V_{so}=13.5$ MeV,\newline $R=r_0\; A^{1/3}_{core}, \; r_0=1.27$ fm, $a=0.7$ fm. These parameters reproduce approximately the experimental energies of the single particle states in $^{133}Sn$ \cite{96prlHoff}. The bound and the resonant states were calculated by using the computer codes \cite{82cpcVertse,95cpcIxaru}. The scattering states were calculated by using Ixaru's constant perturbation method \cite{Ixbook}.

The bound states for valence orbits are defined by the $2p1f0h$ major shell with
the following energies $1f_{7/2}=-2.442$ MeV, $2p_{3/2}=-1.395$ MeV, $0h_{9/2}=-0.923$ MeV, $2p_{1/2}=-0.736$ MeV and $1f_{5/2}=-0.148$ MeV (in this potential the  $0i_{13/2}$ intruder state is unbound). 

\begin{figure}[ht]
\begin{center}
\vspace{8mm}
\includegraphics[width=0.65\textwidth]{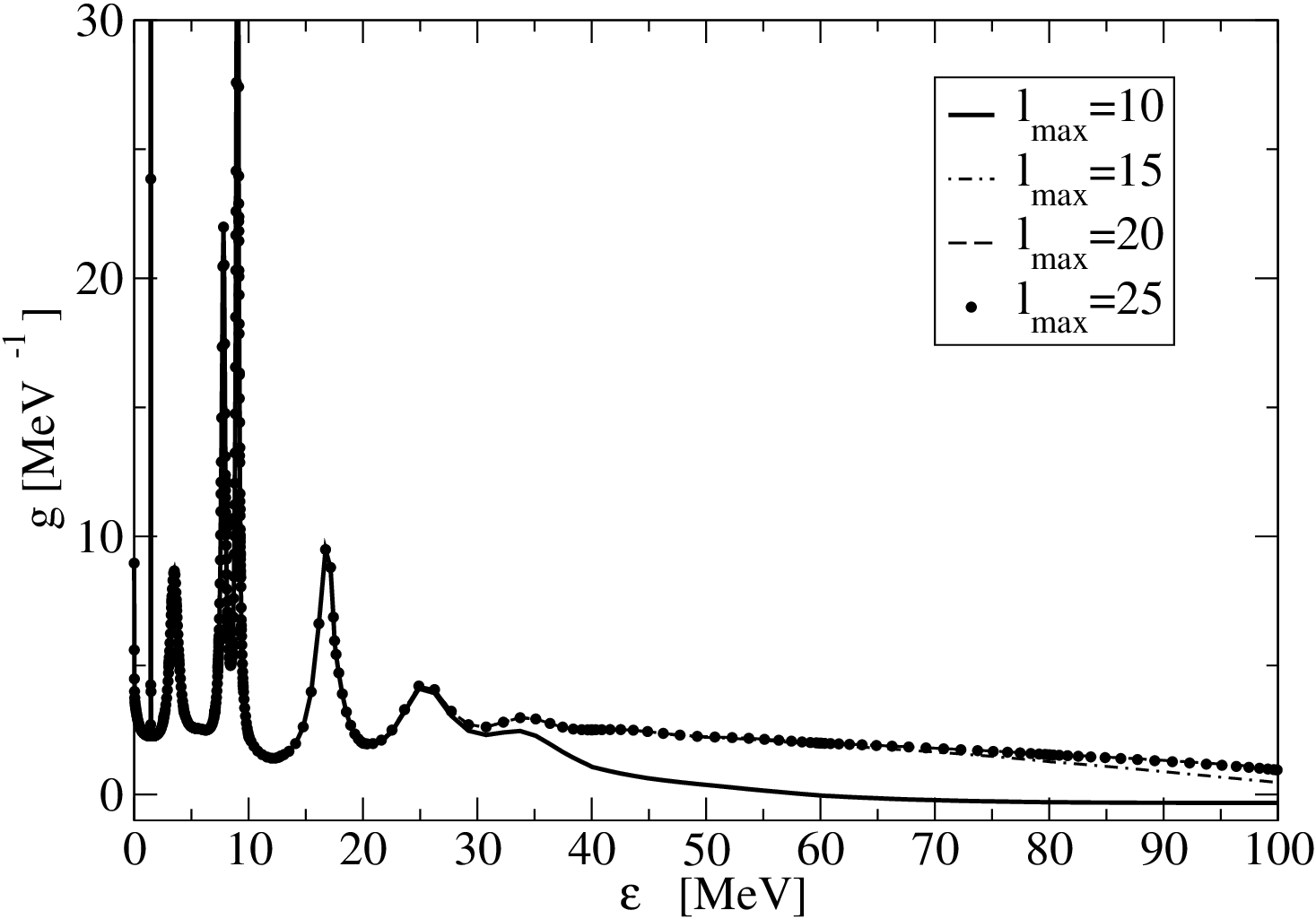}
\caption{\label{fig.ge} CSPLD up to $\varepsilon_{max}=100$ MeV for different $l_{max}$ values.}
\end{center}
\end{figure}

Fig. \ref{fig.ge} shows the $\varepsilon_{max}$ dependence of the CSPLD for different $l_{max}$ values up to $\varepsilon_{max}=100$ MeV. We can easily identify resonances below $20$ MeV. They correspond  to the $i_{13/2}$, $g_{9/2}$, $i_{11/2}$, $k_{15/2}$ and $l_{17/2}$ partial waves (in increasing energy order). We can calculate the widths of the resonances by using the program Gamow \cite{82cpcVertse} and get the widths
$\Gamma(i_{13/2})=0.436\times 10^{-4}$ MeV, 
$\Gamma(g_{9/2})=0.887$ MeV,
$\Gamma(i_{11/2})=0.382$ MeV,
$\Gamma(k_{15/2})=0.251$ MeV,
$\Gamma(l_{17/2})=1.735$ MeV. The $530$ mesh points in energy used for the calculation of the CSPLD were  distributed as  follows: $50$ mesh points were taken between zero and the first resonance. Then $10$ mesh points were taken for each  resonance. Above $20$ MeV  one mesh point/MeV was used.

An increase of the $l_{max}$ value makes the potential less important and  $\delta_c(\varepsilon) \rightarrow 0$ as $l_c \rightarrow \infty$ for a fixed energy and as $\varepsilon \rightarrow \infty$ for a fixed angular momentum \cite{82Newton}. This implies that increasing $l_{max}$ will affect mostly the tail of $g(\varepsilon)$ and will not affect the low energy spectrum in which we are interested in. Therefore it will be enough to take $l_{max}=20$.

We will consider two model spaces: (i) $M_b$: formed by only the valence bound states, and (ii) $M_c$: formed by all partial waves up to $20$
with a cutoff energy at $100$ MeV.

\subsubsection{Pairing Strength}
We choose the nucleus $^{148}$Sn, with $N=16$ as the reference nucleus to fit the strength parameter $\chi$. For this nucleus there is no experimental data
available therefore theoretical masses were used instead \cite{1988Tachibana,1988Tachibana2}. The pairing energy obtained was $P(16)=0.985\;MeV$. Solving the BCS equation for the model space $M_b$ gave $\chi^{BCS}_b=23.66\;MeV$, while for the model space $M_c$, $\chi^{BCS}_c=14.67\;MeV$. Solving the Richardson equations we obtained $\chi^{Rich}_b=18.4\;MeV$, and $\chi^{Rich}_c=12.25\;MeV$. We observed a much smaller pairing strength value in the Richardson model compared to the BCS model. 

A complete representation should include the continuum part of the energy spectrum. The  absence of the continuum in the model space $M_b$ shows up
 in a larger value of the pairing strength $\chi$.

\subsection{Results}
After fixing the model space and the paring strength we can calculate the quantities introduced in the section \ref{sec.method}. 

\subsubsection{BCS model solution}
\begin{figure}[ht] 
\begin{center}
\vspace{8mm}
 \includegraphics[width=0.65\textwidth]{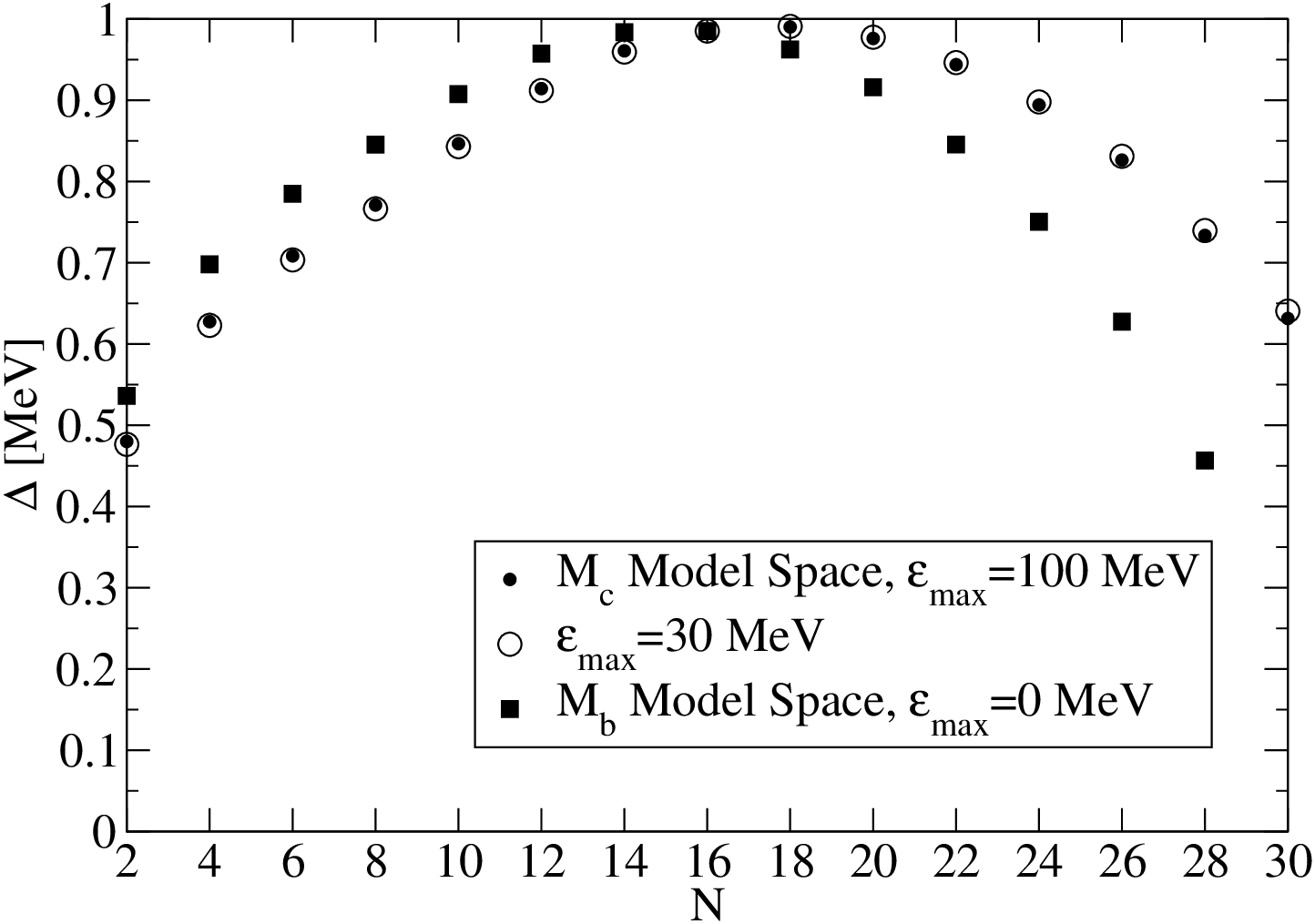}
 \caption{\label{fig.gap} Gap parameter for Tin isotopes in the bound model space ($M_b$), and in two continuum model spaces: $\varepsilon_{max}=100\;MeV$ ($M_c$) and $\varepsilon_{max}=30\;MeV$.}
\end{center}
\end{figure}

The gap parameter $\Delta$ was calculated in both $M_b$ and $M_c$ model spaces for the tin isotopes between $^{134}$Sn and $^{162}$Sn. Their values are shown in fig. \ref{fig.gap} 

In the $M_c$ model space the pairing gap is smaller if $N$ is below $N=16$ and it is larger for $N>16$. As a consequence the isotopes being close to the drip line have a finite gap value, while  with the $M_b$ model space they go to zero rapidly. In order to check the independence of the previous conclusion on the cutoff energy the gap parameters were calculated for a model space with  $\varepsilon_{max}=30\;MeV$ ($\chi=15.58\;MeV$) also  as shown in the same figure.

\begin{figure}[ht] 
\begin{center}
\vspace{8mm}
 \includegraphics[width=0.65\textwidth]{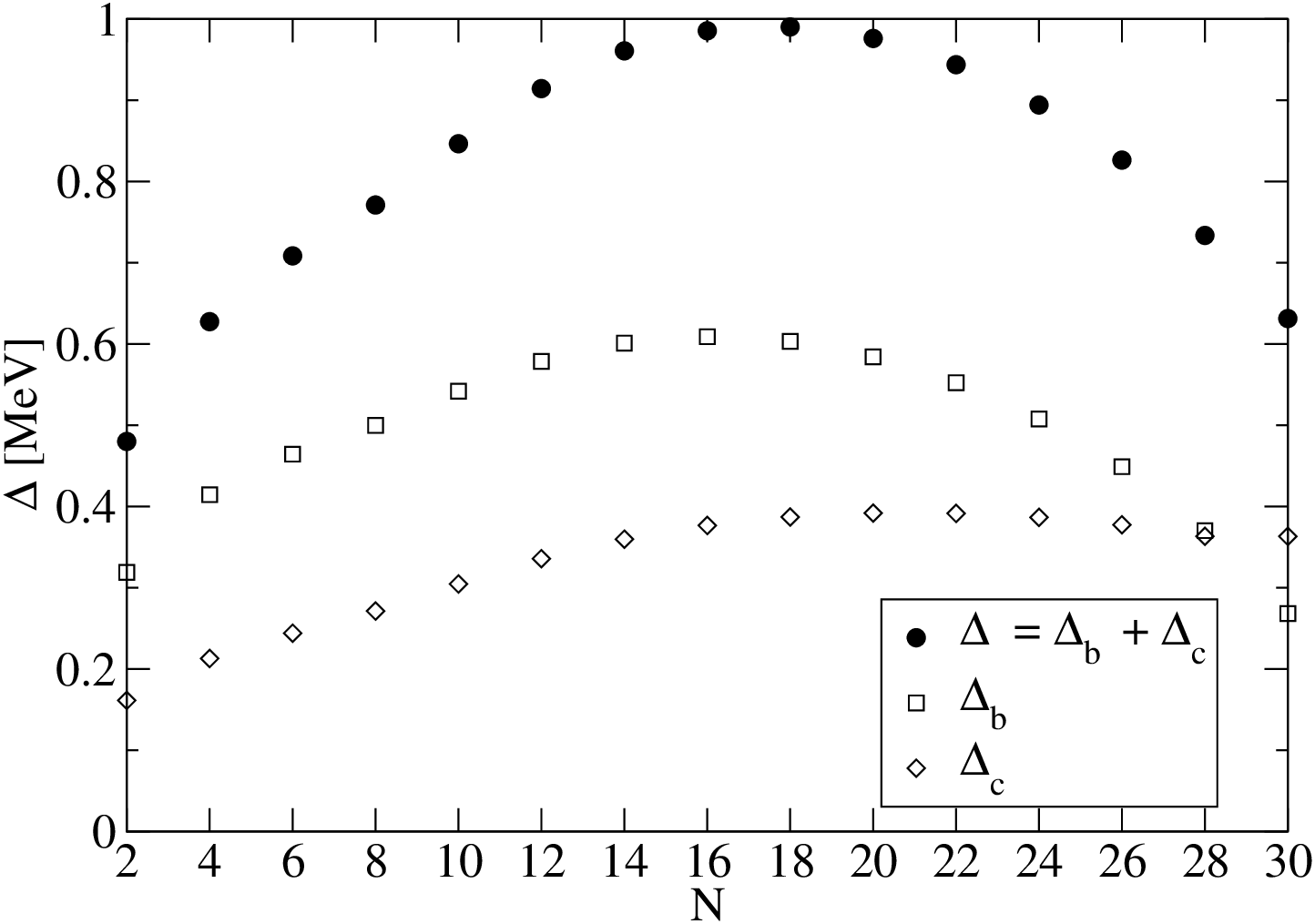}
 \caption{\label{fig.gap_b_c_bc} Relative contribution to the gap parameter for Tin isotopes in the model space $M_c$.}
\end{center}
\end{figure}

Fig. \ref{fig.gap_b_c_bc} shows the contributions of $\Delta_b$ Eq. (\ref{eq.gap-b}) and $\Delta_c$ Eq. (\ref{eq.gap-c}) to the constant gap $\Delta$ in the model space $M_c$. It shows that the continuum part of the spectrum makes an important contribution to the pairing gap.

Because the continuum has an important contribution to the pairing, we can be interested
in the occupation of the states in the continuum. Table \ref{table.n} shows the  contribution of $N_b$ Eq. (\ref{eq.nd}) and $N_c$ Eq. (\ref{eq.nc}) to the particle number $N$. It shows that the ``occupation'' of continuum states does not exceed $4\%$. The same figure was found when we took a different cutoff energy. This a consequence of using the CSPLD relative to the free particle density.
\begin{table}[hb]
\caption{\label{table.n} Contribution from the continuum ($N_c$) and bound ($N_b$) particle number $N=N_c+N_b$ for the model space $M_c$.}
\begin{center}
  \begin{tabular}{ccc}
  N &  $N_b$   &  $N_c$  \\
\hline
 2  &  1.917   &  0.083   \\
 4  &  3.848   &  0.152    \\
 6  &  5.787   &  0.213    \\
 8  &  7.718   &  0.282    \\
 10 &  9.619   &  0.381    \\
 12 &  11.504  &  0.495    \\
 14 &  13.391  &  0.609    \\
 16 &  15.285  &  0.715    \\
 18 &  17.189  &  0.811    \\
 20 &  19.102  &  0.898    \\
 22 &  21.026  &  0.974    \\
 24 &  22.955  &  1.045    \\
 26 &  24.886  &  1.114    \\
 28 &  26.808  &  1.192    \\
 30 &  28.498  &  1.502    \\
  \end{tabular}
 \end{center}
 \end{table}

\begin{figure}[ht] 
\begin{center}
\vspace{8mm}
 \includegraphics[width=0.65\textwidth]{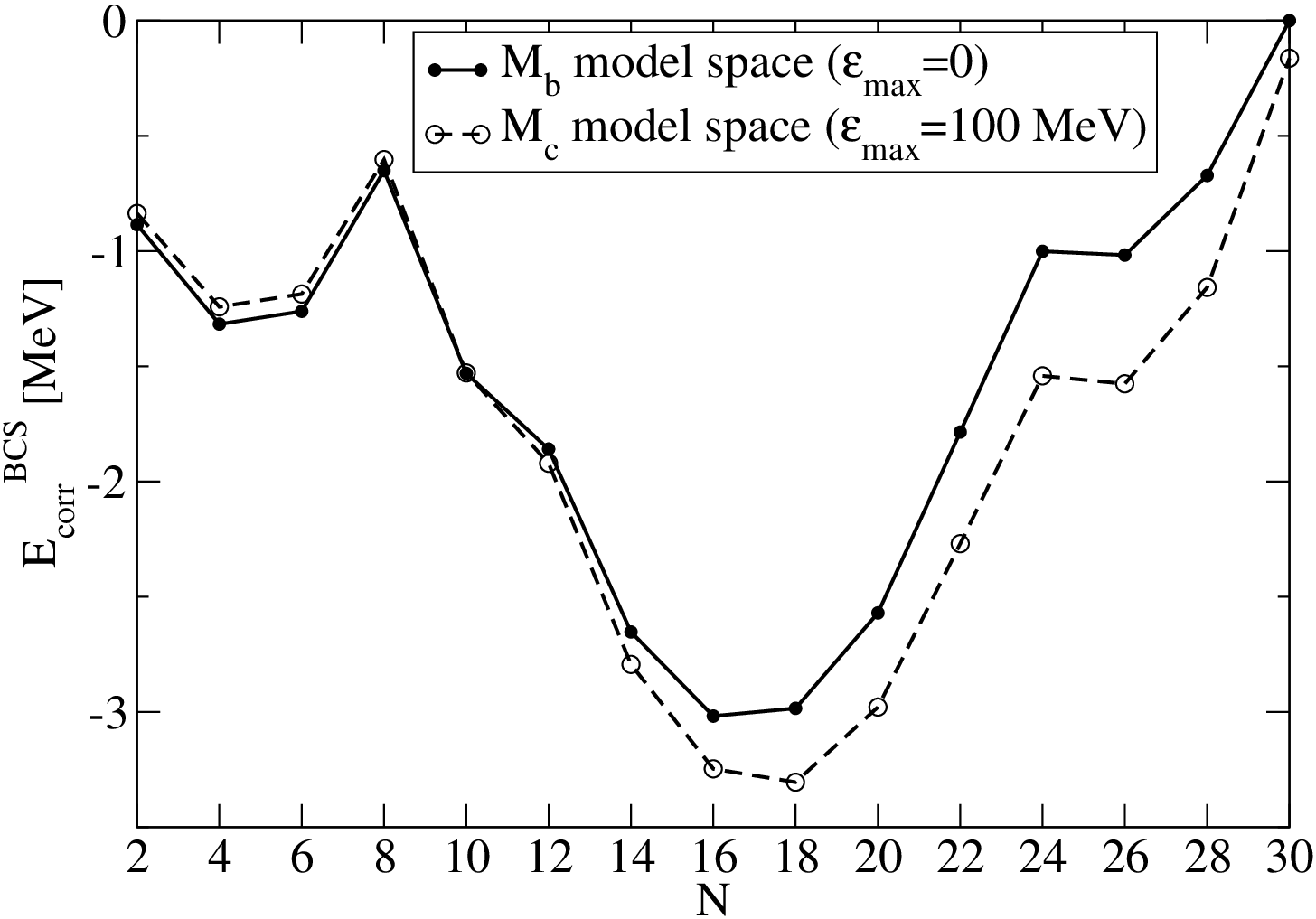}
 \caption{\label{fig.ecorrbcs} Pairing correlated energy in the BCS model solution for the $M_b$ and $M_c$ model spaces.}
\end{center}
\end{figure}

Finally we calculated the correlation energy Eq. (\ref{eq.ecorr}) for the two model spaces $M_b$ and $M_c$ as it shown in figure \ref{fig.ecorrbcs}. It is found that above $^{146}$Sn ($N=14$) the continuum increases the pairing correlation.

\subsubsection{Richardson model solution}
\begin{figure}[ht] 
\begin{center}
\vspace{8mm}
 \includegraphics[width=0.65\textwidth]{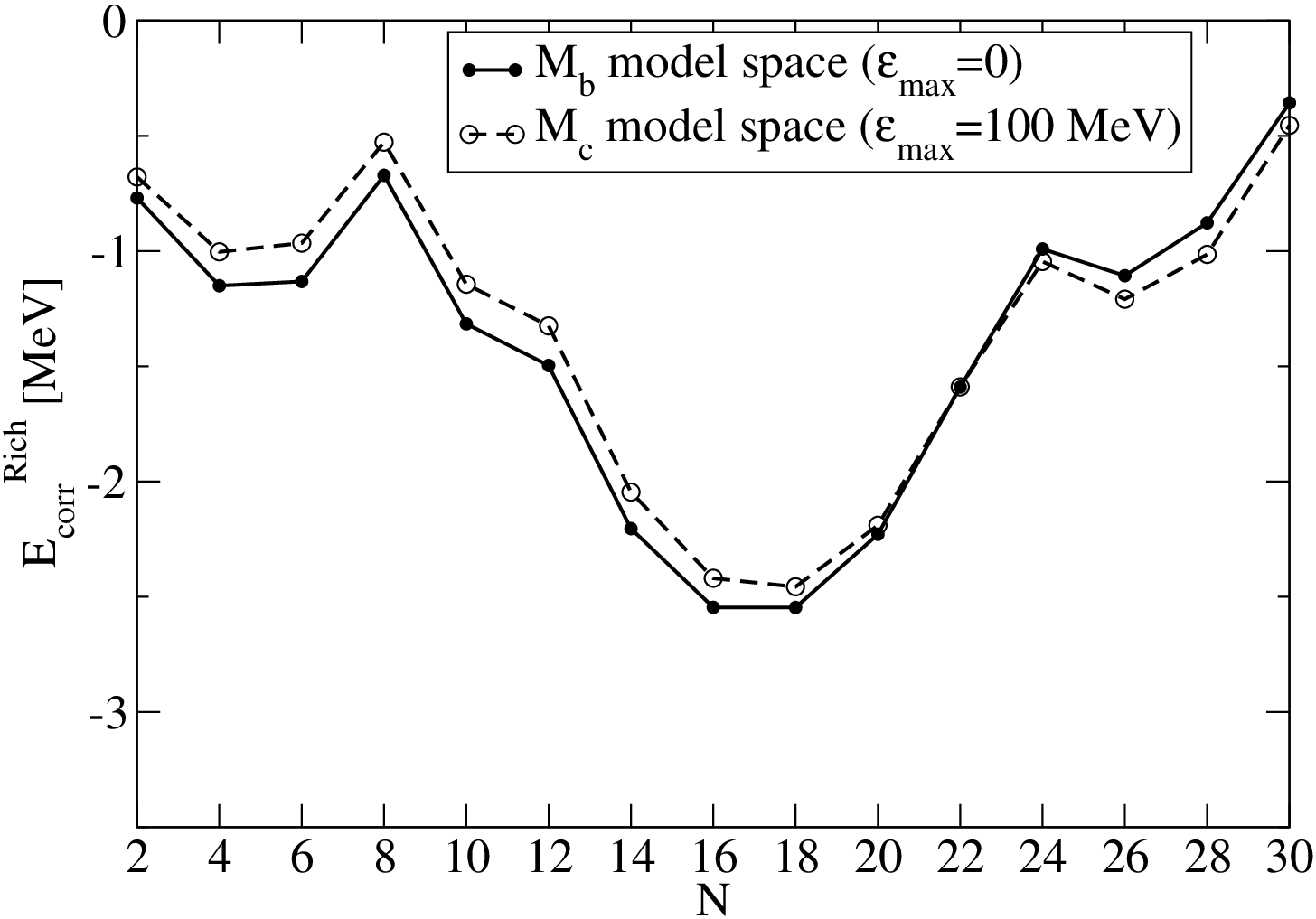}
 \caption{\label{fig.ecorrrich} Pairing correlated energy in the Richardson model solution for the $M_b$ and $M_c$ model spaces.}
\end{center}
\end{figure}

The Richardson model gives the exact solution to the constant pairing Hamiltonian. The pairing correlated energy was calculated using the Richardson model solution in the model space without continuum $M_b$ and the model space with continuum $M_c$. Figure \ref{fig.ecorrrich} shows the result. The solutions in the two models do not depart much from each other.

\subsubsection{Comparison of the solutions in the BCS and Richardson models}
\begin{figure}[ht] 
\begin{center}
\vspace{8mm}
 \includegraphics[width=0.65\textwidth]{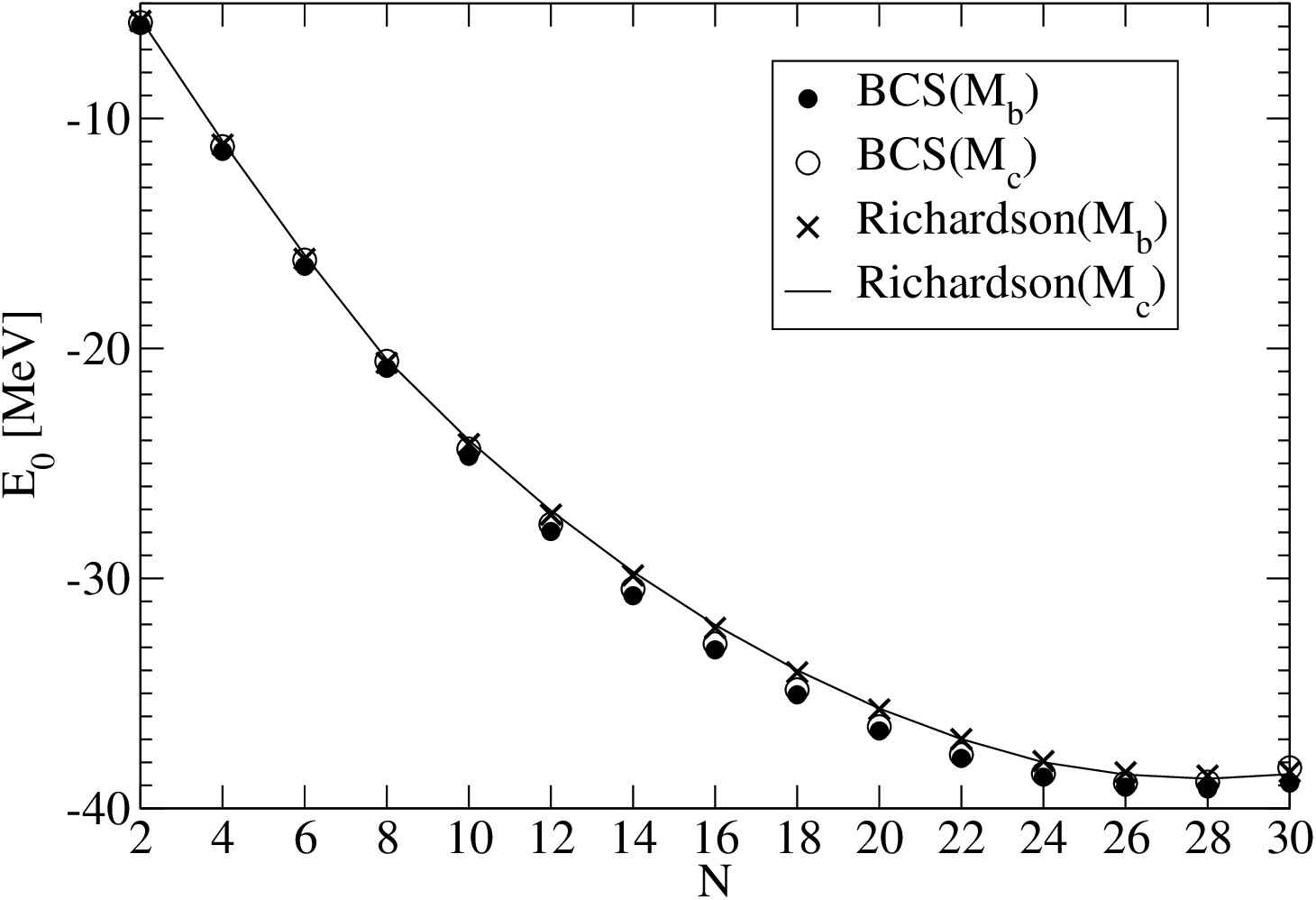}
 \caption{\label{fig.E0vsN_all} Comparison of the ground state energy for the Tin isotopes in the approximated BCS solution and the exact Richardson solution for the two model spaces $M_b$ and $M_c$.}
\end{center}
\end{figure}

The BCS model solution is an approximate solution of the pairing Hamiltonian which does not conserve the particle number, while the  solution in the Richardson model is the exact for the constant pairing Hamiltonian. 
In spite of this huge difference between the two models the ground state energies
for the Tin isotopes $^{134}$Sn-$^{162}$Sn fig. \ref{fig.E0vsN_all}, shows small
differences between the results of the two approaches for the isotopes there.  A small difference appears between the BCS and Richardson solutions after the single particle ground state is filled completely. A small difference between the two model spaces in the BCS solution can be observed from $N=12$ on. The variational BCS solution gives a smaller  energy than  the exact Richardson solution. Since the BCS is a variational solution its energy should be greater than the exact one. This apparent contradiction is resolved if we remember 
that the solutions were calculated with different value of the pairing strength. The clue of the good agreement is that the binding energy is an observable and it should not depend neither on the model solution nor on the model space. The differences between the two model solutions and between the model spaces were absorbed by a single parameter, i.e. the strength parameter $\chi$. Even when this last statement is a trivial affirmation from a theoretical point of view 
it is not trivial in practical calculations.

In the BCS model calculation the pairing-correlated energy shows an enhancement of the pair correlation as the Fermi level $\lambda$ approaches the threshold.
The Richardson model solution does not show a similar trend if we change from the $M_b$ model space to the $M_c$ model space. This shows that  when the continuum is included the BCS approach artificially enhances the pair correlations for a Fermi level being less or around $1$ MeV (for $N=16$ the Fermi level is $\lambda=-1.12$ MeV).

The inclusion of the continuum in the BCS solution predicts that the pairing energy will not be zero at the end of the shell as it happens in a bound representation. The Richardson solution confirms this result with a larger value than the one predicted by the BCS.

By comparing the paring-correlated energies of the two models we found that the BCS overestimates the pair correlation for every isotopes of the chain (except the last one) by an average amount of $278$ keV. The largest difference is $848$ keV at the middle of the isotope chain.

\section{Conclusion} \label{sec.discussion}
In this paper we calculated the many-body constant-pairing model Hamiltonian in the framework of the approximate BCS solution and in the exact Richardson solution. A representation which included the continuum part of the spectrum 
using the level density was used. The level density was defined with respect to the box normalization in order to subtract from the total density the contribution coming from the Fermi gas. We have found that even though the continuum makes an important contribution to the pairing gap, the population of the continuum is weak and its value is independent of the cutoff energy. 

At the scale of the binding energy we have shown that the ground state energy is almost independent of the models considered and it is also indepent on the size of the model space. 
But at the scale of the pair correlated energy the BCS solution does depend on the size of the model space. 
This dependence is enhanced for Fermi energies being around $1$ MeV  below the threshold where the continuum starts. 
The correlated energy is overestimated by an average amount of $300$ keV in the
 approximation which does not conserve the particle number. 
We demonstrated that with the exact solution the results are almost independent of the size of the model space. This poses the question why to include the continuum part of the spectrum into the basis. This question can not be answered with the observables calculated in this paper. It is worth to make an observation at this point, we should include the continuum part of the spectrum 
into the basis because doing so the effective parameter dependence (the paring strength dependence in the constant pairing case) is reduced.
 
The future perspective includes the study of the spectra of the even Carbon isotopes in the Richardson model solution with continuum single particle level density and the comparison of the real energy representation with the complex energy representation of the continuum spectrum.

This work has been partially supported by the National Council of Research PIP-77 (CONICET, Argentina).

\end{document}